\documentclass[onecolumn]{aastex701}

\usepackage{amsmath}
\usepackage{amsfonts}
\usepackage{graphicx,color}
\usepackage{comment}
\usepackage{amsmath}
\usepackage{soul}

\begin{document}

\title{A Comoving Framework for Planet Migration}

\author[0000-0002-2450-7389]{Ximena S. Ramos}
\email[show]{ximena.ramos@umayor.cl}
\affiliation{Centro Multidisciplinario de Física, Vicerrectoría de Investigación, Universidad Mayor, 8580745 Santiago, Chile}

\author[0000-0002-3728-3329]{Pablo Ben\'itez-Llambay}
\email[show]{pablo.benitez@uai.cl}
\affiliation{Facultad de Ingenier\'ia y Ciencias, Universidad Adolfo Ib\'añez, Av. Diagonal las Torres 2640, Peñalol\'en, Chile}

\begin{abstract}
The migration of planets within their nascent protoplanetary disks is a fundamental process that shapes the final architecture of planetary systems. However, studying this phenomenon through direct hydrodynamical simulations is computationally demanding, with traditional methods on fixed grids being ill-suited for tracking planet migration over long timescales due to their high cost and limited domain. In this work, we present a self-consistent comoving framework designed to overcome these challenges. Our method employs a coordinate transformation that scales with the planet's evolving semi-major axis, keeping the planet stationary with respect to its local computational grid. This transforms the standard hydrodynamic equations by introducing a source term that accounts for the inertial forces of the non-inertial reference frame. We implement this framework in the FARGO3D code and validate it through a benchmark test, demonstrating excellent agreement with conventional fixed-grid simulations until the latter are compromised by boundary effects. Our analysis shows that, for long-range migration scenarios, the comoving method can be over an order of magnitude more computationally efficient, dramatically reducing the cost of simulating migrating planets and making secular timescale simulations computationally feasible. This framework serves as both a powerful numerical and theoretical tool, simplifying the time-dependent flow around a migrating planet that offers clearer physical insight. It enables long-term, self-consistent studies of planet-disk interaction, representing a crucial step towards performing planet-population synthesis based on full hydrodynamical simulations.
\end{abstract}

\section{Introduction}

The vast diversity of observed exoplanetary systems indicates that planet migration is a fundamental process in shaping their final architectures \citep[see e.g.,][and references therein]{Armitage2024}. This migration, driven by the gravitational interaction between a planet and its nascent protoplanetary disk, can significantly alter a planet's orbit \citep{Goldreich1980}. The dynamics become even more complex when multiple planets interact simultaneously within the disk, leading to intricate orbital evolutions \citep[see e.g.,][]{Cresswell2006,AtaieeKley2020,Kanagawa2020}.

The outcome of migration exhibits a strong sensitivity to a wide range of physical parameters. The final orbital configuration depends on the planet's mass, the disk's temperature and density gradients, its thermodynamic properties, and the presence of other gravitating bodies, among other factors \citep[see e.g.,][]{Baruteau2014, Paardekooper2023, BenitezLlambay2024}. Despite significant progress in theoretically and numerically characterizing these dependencies, predicting the migration rate and its long-term consequences for a specific planet-disk system remains a formidable challenge.

Current methodologies for studying planet migration have significant limitations. Standard linear torque analyses \citep[e.g.,][]{2002ApJ...565.1257T}, while providing valuable physical insight, often rely on simplified disk structures and, crucially, neglect the planet’s own equation of motion. On the other hand, numerical simulations, though capable of capturing the complex non-linear dynamics, are computationally intensive. To manage this high computational cost, it is customary to adopt compromises, such as fixing the planet on a static orbit to measure torques or restricting simulations to a narrow computational domain where migration is tracked for only a short period. 

These simplifications are not merely inconvenient; they are often fundamentally flawed. Fixing the planet's orbit violates the conservation of total energy and angular momentum for the star-planet-disk system. An important attempt to address this was the remapping method developed by \cite{BenitezLlambay2016b}, which uses a mesh that follows the migrating planet. While this technique allows for long-timescale simulations, in its original form it is not conservative, as it relies on non-conservative linear interpolation to populate the new mesh at each step. This repeated interpolation can introduce numerical diffusion that affects the long-term evolution of the system.

To overcome these limitations, we introduce in this paper a novel framework built upon a comoving coordinate system designed for the self-consistent study of planet migration. The core of our approach is to solve the hydrodynamic equations on a computational grid that is centered on the planet and whose spatial scale is defined by the planet's instantaneous semi-major axis. A similar strategy was previously employed by \cite{2006PhDT.........3P}, who exploited the covariant formalism specific to the RODEO code to keep the planet fixed within the mesh. Here, we develop a framework that can be applied to any numerical hydrodynamics code by simply adding the appropriate source terms to the standard equations of motion.
This strategy directly addresses the primary challenges of traditional methods. By keeping the planet stationary with respect to the grid, our framework obviates the need for an extended computational domain that must encompass the entire migration path. Consequently, the numerical cost of simulating a migrating planet is dramatically reduced, becoming comparable to that of simulating a static planet on a domain of the same relative size. This efficiency enables long-term, self-consistent simulations of planet-disk interaction. The framework is defined by a radial coordinate transformation that is independent of the frame's rotation. As such, it can be readily combined with standard orbital advection algorithms, requiring only minimal modifications to existing hydrodynamics codes.

This paper is organized as follows. In Section \ref{sec:basic_equations}, we review the fundamental equations of the planet-disk system. Section \ref{sec:coordinate_transformation} introduces the comoving coordinate transformation and presents the derivation of the resulting fluid equations. The practical application of our framework is detailed in Section \ref{sec:implementation}, where we derive a first-order estimate for the computational speed-up and describe the implementation within the FARGO3D code. In Section \ref{sec:benchmark}, we validate our method through a benchmark test, comparing it against a traditional fixed-grid simulation to demonstrate its accuracy and advantages. We discuss the broader implications and potential future applications of our work in Section \ref{sec:discussion}. Finally, we summarize our results and present our conclusions in Section \ref{sec:summary_conclusion}.

\section{Basic equations}
\label{sec:basic_equations}

We model a single planet of mass $m_{\rm p}$ embedded in a locally isothermal, gaseous protoplanetary disk that orbits a central star of mass $M_\star$. The planet-disk system is described by the continuity and momentum equations for the gas, coupled with the equation of motion for the planet.

The evolution of the gas is governed by the continuity equation,
\begin{equation}
\label{eq:continuity}
\partial_t \rho + \nabla\cdot\left(\rho {\bf v}\right) = 0\,,
\end{equation}
and the momentum equation,
\begin{equation}
\label{eq:momentum}
\partial_t {\bf v} + \left({\bf v} \cdot \nabla\right) {\bf v} = -\frac{\nabla P}{\rho} -\nabla \phi + {\bf A}\,.
\end{equation}
Here, $\rho$, $\mathbf{v}$, and $P$ are the gas density, velocity, and pressure, respectively. The term ${\bf A}$ on the right-hand side of Equation~\eqref{eq:momentum} accounts for additional accelerations, such as the one produced by viscosity. 
The gravitational potential in an astrocentric frame, $\phi$, is given by
\begin{equation}
\label{eq:potential}
\phi = - \frac{GM_\star}{|{\bf r}|} - \frac{Gm_{\rm p}}{|{\bf r}-{\bf r}_{\rm p}|} + \frac{G m_{\rm p} {\bf r} \cdot{\bf r}_{\rm p}}{{|{\bf r}_{\rm p}}|^3} + {\bf r} \cdot G\int \rho \frac{{\bf r}}{|{\bf r}|^3} \, d V\,.
\end{equation}
Here, ${\bf r}$ and ${\bf r}_{\rm p}$ represent the position vectors of a gas element and the planet relative to the central star. $G$ is the gravitational constant. The first two terms describe the gravitational potential of the central star and the planet. The third and fourth terms account for the indirect potential arising from the reflex motion of the star due to the planet and the disk. The disk contribution is obtained by integrating over all mass elements $dm = \rho\,dV$.

The planet's motion is described by its equation of motion
\begin{equation}
\label{eq:planet}
\frac{d^2 {\bf r}_{\rm p}}{dt^2} = -\Omega_{\rm p}^2 {\bf r}_{\rm p}  + G \int \rho \frac{{\bf r}-{\bf r}_{\rm p}}{{|{\bf r} -{\bf r}_{\rm p}|}^3} \, d V  -q\Omega_{\rm p}^2 {\bf r}_{\rm p} - G\int \rho \frac{{\bf r}}{|{\bf r}|^3} \, d V\,.
\end{equation}
The first and second terms on the right-hand side represent the gravitational forces exerted by the central star and the disk, respectively. The third and fourth terms account for the indirect forces arising from the acceleration of the central star induced by the planet and the disk.

\subsection{Initial and boundary conditions}

To solve the system of equations \eqref{eq:continuity}-\eqref{eq:planet}, we must prescribe initial and boundary conditions that establish a well-defined, near-equilibrium state. A typical setup begins with the planet on a circular orbit at $r=a_{\rm p}$ within a rotationally-supported gaseous disk. In two-dimensions, this initial state is constructed by defining a surface density profile, $\Sigma(r)$, which is commonly a smooth power-law, and a temperature profile, $T(r)$, which for a locally isothermal disk is often set also as a power-law. These properties in turn determine the initial gas velocity field, ${\bf v}$, where the azimuthal velocity is slightly sub-Keplerian to account for radial pressure support. For three-dimensional models, vertical hydrostatic equilibrium must also be established, for which models exist based on, for example, vertically isothermal layers at constant cylindrical or spherical radii \citep{Nelson2013, Masset2016}. In addition to this initial state, boundary conditions are applied at the edges of the computational domain to enforce a physically realistic behavior. At or near the domain edges, inflow/outflow or zero-gradient conditions manage the flow of gas (or solids) and are often supplemented by wave-damping zones to prevent wave reflections \citep{deValBorro2006}. 
A crucial limitation of this standard approach is that the radial boundaries are defined at fixed locations. This static grid presents a key challenge for long-term simulations, as a migrating planet will eventually approach a boundary, compromising the results. This fundamental problem motivates the development of the comoving framework detailed in the next section. Hereafter, we will refer to calculations performed on this standard, fixed grid as being in the inertial or rest frame. While an astrocentric frame is not strictly inertial, this terminology serves to create a clear contrast with the explicitly highly non-inertial, accelerating nature of the comoving frame.

\section{Comoving frame}
\label{sec:coordinate_transformation}

We define a comoving coordinate system that enables solving hydrodynamic equations by tracking planet migration on a fixed computational grid. In this system, the planet's semi-major axis remains constant relative to the mesh coordinates. This choice maintains a fixed representation of the planet's orbit within the comoving frame, although our framework is adaptable to alternative definitions based on other time-dependent orbital parameters, such as the instantaneous radial distance or any other radial reference point. It is worth noticing that the framework presented here can be applied to the case of multiple planets, which is discussed in section \ref{sec:discussion}.

\subsection{Comoving transformation}
\label{sec:comoving_transformation}

The core of our comoving framework is a coordinate transformation that rescales space and time to follow the migrating planet, keeping its semi-major axis constant relative to the mesh coordinates. To achieve this, we redefine our fundamental unit of length to be the planet's instantaneous semi-major axis, $a_{\rm p}(t)$. By measuring all radial distances in units of $a_{\rm p}$, a planet on a circular orbit remains at a constant dimensionless radius of unity, even as its physical distance from the star changes. Similarly, we rescale time according to the planet's Keplerian orbital frequency, $\Omega_{\rm p}(t)  = \sqrt{GM_\star/a_{\rm p}^3}$, which represents the natural timescale of the system at the planet's location. These two transformations imply locally that $d{\bf r}' = d{\bf r}/a_{\rm p}$ and $dt' = \Omega_{\rm p} dt$, being ${\bf r}'$ and $t'$ the comoving position vector and time coordinates. These definitions can be expressed as the following coordinate transformation
\begin{align}
\label{eq:tprime}
t' &= \int_{t_0}^t \Omega_{\rm p}(s) ds\,, \\
\label{eq:rprime}
{\bf r}' &= \frac{\bf r}{a_{\rm p}}\,.
\end{align}
Under this transformation, derivatives in old variables are written in terms of the new ones as
\begin{align}
\partial_t &= \Omega_p \left[\partial_{t'} -\mathcal{H} \left(\mathbf{r}' \cdot \nabla'\right)\right]\,,
\label{eq:transform_time}\\
\nabla &= a_{\rm p}^{-1} \nabla'\,, 
\label{eq:transform_space}
\end{align}
with 
\begin{equation}
    \label{eq:migration-rate}
    \mathcal{H} = \frac{\dot{a}_{\rm p}}{v_{\rm p}} = \frac{1}{a_{\rm p}}\frac{da_{\rm p}}{dt'}\,,
\end{equation} 
where $\dot{a}_{\rm p} = da_{\rm p}/dt$ and $v_{\rm p} = a_{\rm p}\Omega_{\rm p}$. Eq.\,\eqref{eq:migration-rate} corresponds to the dimensionless scaling rate, which is equal to the inverse of the comoving migration timescale.  We define the following dimensionless comoving quantities
\begin{align}
\label{eq:comoving_density}
{\rho}'  &=  M_{\rm \star}^{-1} a_{\rm p}^{d} \rho \,, \\
\label{eq:comoving_velocity2}
{\bf v}' &= v_{\rm p}^{-1} {\bf v}\,, \\
\label{eq:comoving_potential}
\phi'  &= v_{\rm p}^{-2} \phi\,, \\
\label{eq:comoving_pressure}
P' &= M_{\rm \star}^{-1}a_{\rm p}^{d}  v_{\rm p}^{-2} P\,,
\end{align}
where $d$ is the dimensionality of the system ($d=1, 2, \text{or } 3$). Under these definitions, the disk potential $\phi'$ becomes
\begin{equation}
\label{eq:potential_comoving}
\phi' = -\frac{1}{|{\bf r}'|} - \frac{q}{|{\bf r}'-{\bf r}'_{\rm p}|} + \frac{q {\bf r}'\cdot{\bf r}'_{\rm p}}{|{\bf r}'_{\rm p}|^3} + {\bf r}' \cdot \int \rho' \frac{{\bf r}'}{{|{\bf r}'|}^3} \, d V' \,,
\end{equation}
where $q = m_{\rm p}/M_\star$ is the planet-to-star mass ratio and $dV' = dV/a_{\rm p}^d$. The vector ${\bf r}_{\rm p}'$ is the position vector of the planet, normalized by $a_{\rm p}$.

After defining the dimensionless comoving velocity vector
\begin{equation}
\label{eq:comoving_velocity}
    {\bf u}' = {\bf v}' - \mathcal{H} {\bf r}'\,,
\end{equation}
Eqs. \eqref{eq:continuity}-\eqref{eq:momentum} become
\begin{align}
\label{eq:continuity_prime}
\partial_{t'} {\rho}' + \nabla' \cdot\left({\rho}' {\bf u}'\right) &= 0 \,, \\
\label{eq:momentum_prime}
\partial_{t'} {\bf u}' + \left({\bf u}' \cdot \nabla'\right) {\bf u}' &= -\frac{\nabla' {P}'}{{\rho}'} -\nabla' {\phi}' + {\bf A}' + {\bf S}'\,,
\end{align}
where ${\bf A}' = {\bf A}/(v_{\rm p}\Omega_{\rm p})$ and the source term ${\bf S}'$ is given by
\begin{align}
\label{eq:source_term}
{\bf S}' = \left(\frac{1}{2}\mathcal{H}^2 - \frac{d\mathcal{H}}{dt'}\right) {\bf r}' - \frac{\mathcal{H}}{2} {\bf u}'. 
\end{align}

Eqs.\,\eqref{eq:continuity_prime} and \eqref{eq:momentum_prime} are identical to their version in the inertial frame except from the additional inertial source term ${\bf S}'$. 
For clarity in the derivation, the equations presented here are in a non-rotating reference frame. However, in the case of a corotating frame, additional centrifugal and Coriolis terms must be added to the momentum equation \eqref{eq:momentum_prime}. These terms are separable from the new source term ${\bf S}'$, which exclusively handles the effects of the radial scaling.

The new source term \eqref{eq:source_term} encapsulates the inertial forces arising from the transformation to a non-inertial reference frame that radially expands or contracts as a consequence of planet migration. A detailed analysis of its components reveals their distinct physical nature. 
The first term of \eqref{eq:source_term}, which depends on the position vector ${\bf r}'$, arises from the inertial acceleration of the coordinate grid itself. It consists of two contributions: a term proportional to the time derivative of the expansion rate, $-d\mathcal{H}/dt'$, which accounts for the non-uniform acceleration of the frame's scaling, and a centrifugal-like term, $\mathcal{H}^2/2$, which produces a purely radial, outward-directed acceleration independent of the fluid velocity.  
The second term constitutes an isotropic damping/anti-damping term that depends linearly on the peculiar velocity. Its physical origin lies in the time-evolution of the system's fundamental velocity scale, $v_{\rm p}$. Because the unit of velocity scales as $a_{\rm p}^{-1/2}$, the dimensionless velocity ${\bf u}'$ must evolve to compensate for this changing metric, ensuring that the simulation remains consistent with the conservation laws of the inertial frame. 
The comoving term $-\mathcal{H}/2{\bf u}'$ acts as the specific fictitious force that drives this required acceleration in the dimensionless variables, effectively enforcing the physical consistency of the system within the non-conservative comoving framework.

\subsection{Viscosity}
\label{sec:viscosity}

In protoplanetary disks studies it is typical to consider the effect of viscosity as a source term ${\bf A}$. Here, we show how to treat this term within the comoving framework. 
Viscosity is added to Eq.\,\eqref{eq:momentum} through the divergence of the viscous stress tensor as
\begin{equation}
\label{eq:viscous_term}
    {\bf A} = \frac{1}{\rho}\nabla \cdot {\bf T}\,,
\end{equation}
with 
\begin{equation}
    {\bf T} = \rho \nu \left[ \nabla {\bf v} + (\nabla {\bf v})^{\rm T} - \lambda (\nabla \cdot {\bf v}) {\bf I} \right]\,,
\end{equation}
where {\bf I} is the identity tensor and $\lambda = 2/d$ is a coefficient chosen such that the bulk viscosity is zero.
In the $\alpha$-disk formalism \citep{Shakura1973}, the kinematic viscosity can be written as $\nu = \alpha_{\nu} h^2 v_{\rm k} r$, with $h$ the disk aspect-ratio, $v_{\rm k}$ the Keplerian velocity, and $\alpha_{\nu}$ the constant kinematic viscosity parameter. 
Each term of the stress-tensor can be transformed to comoving coordinates, obtaining
\begin{equation}
    \nabla {\bf v} = \Omega_{\rm p}\left( \nabla'{\bf u'} + \mathcal{H} {\bf I} \right)\,,
\end{equation}
\begin{equation}
    \nabla \cdot {\bf v} = \Omega_{\rm p} \left( \nabla' \cdot {\bf u'} + d \mathcal{H} \right)\,.
\end{equation}
Therefore, in comoving coordinates, the viscous stress tensor becomes
\begin{align}
    {\bf T} = T_{\rm p} {\bf T}'\,,
\end{align}
with
\begin{equation}
\label{eq:viscous_comoving}
 {\bf T}' = {\rho}' {\nu}' \left[ \nabla' {\bf u}' + (\nabla' {\bf u}')^{\rm T} - \lambda (\nabla' \cdot {\bf u}') {\bf I} + \left(2-\lambda d\right) \mathcal{H} {\bf I} \right]\,,
\end{equation}
and
\begin{equation}
    T_{\rm p} =  \frac{M_\star}{a_{\rm p}^d}  v_{\rm p}^2\,,
\end{equation}
\begin{equation}
    {\nu}' = \alpha_{\nu} h^2 {v}_{\rm k}' r'\,.
\end{equation}
Close inspection of Eq.\,\eqref{eq:viscous_comoving} reveals that the condition $\lambda = 2/d$ ensures invariance of the viscous stress tensor, which coincides with the zero bulk viscosity hypothesis.
Finally, for the system to admit a steady-state solution, the dimensionless viscosity coefficient $\nu'$ must also remain constant as the frame scales. This strictly requires a physical viscosity profile scaling as $\nu \propto r^{1/2}$ (standard non-flared $\alpha$-disk). For general viscosity laws, the equations retain their standard form, but the coefficients become time-dependent.

Finally, Eq.\,\eqref{eq:viscous_term} in the $\alpha$-disk formalism becomes
\begin{equation}
\label{eq:viscous_term_comoving}
{\bf A} = v_{\rm p} \Omega_{\rm p} \left(\frac{1}{{\rho}'} \nabla' \cdot {\bf {T}'}\right)\,.
\end{equation}

\subsection{Gas and Dust Drag}
\label{sec:gas_dust}

When considering a mixture of gas and dust, the momentum exchange between the two species is included as a source term ${\bf A}$ in their respective momentum equations. For a local dust-to-gas mass ratio, $\varepsilon$, the drag accelerations are characterized by the dimensionless Stokes number, $S_t$. The specific collision terms are given by \citep[see e.g.,][]{BenitezLlambay2019}
\begin{align}
    {\bf A}_{\rm gas} &= -\frac{\varepsilon \Omega_{\rm K}}{{S}_t} \left({\bf v}_{\rm gas} - {\bf v}_{\rm dust}\right)\,, \\
    {\bf A}_{\rm dust} &= -\frac{\Omega_{\rm K}}{{ S}_t} \left({\bf v}_{\rm dust} - {\bf v}_{\rm gas}\right)\,.
\end{align}
When expressed in dimensionless comoving variables, the form of these drag terms remains invariant. The transformation from the physical acceleration ${\bf A}$ to the dimensionless comoving acceleration ${\bf A}'$ follows the scaling established previously
\begin{equation}
    {\bf A}_{\rm gas/\rm dust} = v_{\rm p}\Omega_{\rm p}{\bf A}'_{\rm gas/\rm dust}\,.
\end{equation}
Furthermore, a significant simplification arises in the comoving frame. Because the drag acceleration depends only on the relative velocity between the two species, the term $\mathcal{H} {\bf r}'$ arising from the frame's expansion cancels out. Therefore, the drag can be calculated equivalently using either the velocity ${\bf v}'$ or the comoving velocity ${\bf u}'$.

\subsection{Magnetohydrodynamics}
\label{sec:mhd}

The inclusion of MHD into the models requires adding the induction equation
\begin{equation}
\label{eq:induction}
    \partial_t {\bf B} = \nabla \times \left( {\bf v}\times {\bf B}\right)\,,
\end{equation}
and the Lorentz force in the R.H.S of Eq.\,\eqref{eq:momentum}\,,
\begin{equation}
\label{eq:lorentz}
    {\bf A}_{\rm LF} = \frac{1}{\rho \mu_0} \left(\nabla \times {\bf B}\right) \times {\bf B}\,,
\end{equation}
with \textbf{ B} is the magnetic field and $\mu_0$ the vacuum permeability.

In comoving coordinates, the induction equation becomes

\begin{equation}
    \partial_{t'} {\bf B} = \nabla' \times \left( {\bf u}' \times {\bf B}\right) +  \mathcal{H} (1-d) {\bf B}\,.
\end{equation}
We require the dimensionless magnetic pressure $P_{mag}\propto B^2$ to scale exactly like the thermal pressure $P$ (see Eq. \ref{eq:comoving_pressure}), which implies
\begin{equation}
    {\bf B}' = M_\star^{-1/2} a_{\rm p}^{d/2} v_{\rm p}^{-1} {\bf B}\,,
\end{equation}
and the induction equation in comoving coordinates becomes
\begin{equation}
\label{eq:induction_comoving}
    \partial_{t'} {\bf B'} = \nabla' \times \left( {\bf u}' \times {\bf B'}\right) +  \frac{3-d}{2} \mathcal{H}  {\bf B'}\,.
\end{equation}
The second term of Eq.\,\eqref{eq:induction_comoving} vanishes if $d=3$, which implies that the induction equation in comoving coordinates is fully invariant only in 3D.

Finally, the Lorentz force in comoving coordinates becomes
\begin{equation}
\label{eq:lorentz_comoving}
    {\bf A}_{\rm LF} = v_{\rm p} \Omega_{\rm p} {\bf A}'_{\rm LF}\,,
\end{equation}
with 
\begin{equation}
    {\bf A}'_{\rm LF} = \frac{1}{\rho'\mu_0} \left(\nabla' \times {\bf B'}\right) \times {\bf B'}\,.
\end{equation}
Therefore, the Lorentz force has the same scaling as all of the other accelerations discussed.

\subsection{Energy equation}
\label{sec:energy_equation}

When the internal energy, $e$, is free to evolve, we must solve the energy equation
\begin{equation}
\label{eq:energy}
\partial_t e + \nabla \cdot (e {\bf v}) = -P \nabla \cdot {\bf v}\, .
\end{equation}

We require the thermal pressure ($\propto e$) to scale according to Eq. \eqref{eq:comoving_pressure}, the comoving energy is defined as
\begin{equation}
 e' = M_\star^{-1} a_{\rm p}^d v_{\rm p}^{-2} e\,,
\end{equation}
and Eq.\,\eqref{eq:energy} in comoving coordinates becomes

\begin{equation}
\label{eq:energy_comoving}
\partial_t e' + \nabla' \cdot (e'{\bf u}') = -P' \nabla \cdot {\bf u}' + \mathcal{H} \left(e' - dP'\right)\,.
\end{equation}
For an ideal gas law, $P = \left(\gamma-1\right) e$, and we get
\begin{equation}
\label{eq:energy_comoving_ideal}
\partial_t e' + \nabla' \cdot (e'{\bf u}') = -P' \nabla \cdot {\bf u}' + \mathcal{H} \left[1 - d(\gamma-1)\right]e'\,,
\end{equation}
which is invariant only if $\gamma = (1+d)/d$. In 3D this is possible when $\gamma=4/3$, in 2D if $\gamma = 3/2$. For general gasses, the energy equation is not invariant as it requires a geometric source term to account for the energy change of the gas due to the scaling of the domain. This source term represents the adiabatic cooling (or heating) required to account for the work done by the expansion (or contraction) of the coordinate metric itself.

\subsection{The case of non-constant $h$}
\label{sec:non_constant_h}

In detailed disk models, the disk aspect-ratio is not constant but rather a function of the radial distance from the star. This flaring of the disk is typically modeled as a power-law of the form
\begin{equation}
    \label{eq:h}
    h = h_0 \left( \frac{r}{r_0} \right)^f,
\end{equation}
where $f$ is the disk's flaring index and $h_0$ is the aspect-ratio at a given reference radius $r_0$. Eq.\,\eqref{eq:h} can be written in terms of $a_{\rm p}$ as
\begin{equation} 
    \label{eq:hp}
    h = h_{\rm p}(t') r'^{f}\,,
\end{equation}
with $h_{\rm p}(t') = h_0 (a_{\rm p}(t')/r_0)^{f}$.
This shows that the aspect-ratio is dependent on the time-varying semi-major axis, $a_p(t')$.
Consequently, if $f\neq0$, physical terms that depend on the aspect-ratio such as the gas pressure and the turbulent viscosity become time-dependent and are no longer invariant. It is important to note that this time evolution of $h$ is not introduced by the comoving transformation itself but it is an intrinsic property of the system, which adds numerical complexities that require special treatment in long-term simulations as the mesh initial resolution may become insufficient as the disks becomes thinner in the inner regions. 
One way of dealing with this issue would be to further compress/expand the mesh as the planet migrates to maintain the resolution per scale-height of the disk. A more interesting possibility is to work on a coordinate transformation of the form $d{\bf r}'=d{\bf r}/H_{\rm p}$ \citep[see e.g.,][]{Ward1997a}. For the case of a non-flared disk, this transformation leads to $r'\propto \log(r/r_0)$, which justifies the standard logarithmic mesh used in planet-disk interaction. For $f\neq 0$, $d{\bf r}'\propto d{\bf r}/r^{f+1}$, giving to a another spacing law that strictly maintains resolution per scale-height. In this paper we adopt the simplifying assumption of a constant aspect-ratio ($f=0$), such that $h=h_0$ throughout the entire domain of interest.

To illustrate explicitly the dependence of $f$ in the resulting fluid equations, we write the pressure in the locally isothermal approximation, for which $P = c_{\rm s}^2 \rho$, with $c_{\rm s} = h v_{\rm K}$ the sound speed. 
The comoving sound speed is
\begin{equation}
    c_{\rm s}'(r', t') \equiv \frac{c_{\rm s}}{v_{\rm p}} = h_0 \left(\frac{a_{\rm p}(t')}{r_0}\right)^f r'^{f-1/2}\,,
\end{equation}
which implies that the comoving pressure is
\begin{equation}
    P'(r',t') = h_0^2 \left(\frac{a_{\rm p}(t')}{r_0}\right)^{2f} r'^{2f-1} \rho'\,.
\end{equation}
Therefore, it is clear that time dependence disappears only if $f=0$, for which
\begin{equation}
    P' = h_0^2 r'^{-1} \rho'.
\end{equation}

\subsection{Planet's equation of motion}
\label{sec:equation_of_motion_planet}

While theoretical considerations might suggest modifications to the planet's equation of motion \eqref{eq:planet} within a comoving framework, there is a crucial practical distinction; 
the hydrodynamical calculations, which are bound by a predefined computational grid, represent the limiting factor in computational complexity. 
In contrast, the planet's trajectory is not similarly constrained. 
Therefore, to achieve the simplest and most computationally efficient implementation of the comoving framework, we solve only the hydrodynamics in comoving coordinates, while the planet(s) continue to be solved in the rest frame. In practical terms, eq.\,\eqref{eq:planet} is solved in the comoving framework as the following first-order differential equations
\begin{align}
    \Omega_{\rm p} \frac{d {\bf r}_{\rm p}}{dt'} &= {\bf v}_{\rm p}\,, \\
    \Omega_{\rm p}\frac{d {\bf v}_{\rm p}}{dt'} &= -\Omega_{\rm p}^2 {\bf r}_{\rm p}  + G M_\star \int \rho' \frac{{\bf r}-{\bf r}_{\rm p}}{{|{\bf r} -{\bf r}_{\rm p}|}^3} \, d V'  -q\Omega_{\rm p}^2 {\bf r}_{\rm p} - G M_\star \int \rho' \frac{{\bf r}}{|{\bf r}|^3} \, d V'\,.
\end{align}

For completeness, we include the equation of motion of the planet in comoving coordinates after applying the coordinate transformation to the comoving frame and neglecting the indirect accelerations
\begin{align}
\frac{d {\bf r}'_{\rm p}}{dt'}  &= {\bf v}'_{\rm p} - \mathcal{H} {\bf r}_{\rm p}'\,, \\
\frac{d {\bf v}'_{\rm p}}{dt'}  &= -{\bf r}'_{\rm p} + \int {\rho}' \frac{{\bf r}'-{\bf r}'_{\rm p}}{{|{\bf r}'-{\bf r}'_{\rm p}|}^3} dV' + \frac{1}{2} \mathcal{H} {\bf v}'_{\rm p}\,,
\end{align}
which can be combined as
\begin{align}
\label{eq:planet_comoving_scalar}
\frac{d}{dt'} \left({r'_{\rm p}}^2 + {v'_{\rm p}}^2\right)  + \mathcal{H} \left( 2{r'_{\rm p}}^2 - {v'_{\rm p}}^2\right) 
=  2{\bf v}_{\rm p}' \cdot \int {\rho}' \frac{{\bf r}'-{\bf r}'_{\rm p}}{{|{\bf r}'-{\bf r}'_{\rm p}|}^3} dV'\,.
\end{align}
When the planet is in steady-state in the comoving frame ($dt'= 0$), $dr'_p/dt' = dv'_p/dt = 0$, $r'_p = v'_p = 1$, and Eq.\,\eqref{eq:planet_comoving_scalar} becomes
\begin{align}
\mathcal{H} 
&=  2{\bf v}_{\rm p}' \cdot \int {\rho}' \frac{{\bf r}'-{\bf r}'_{\rm p}}{{|{\bf r}'-{\bf r}'_{\rm p}|}^3} dV'\,.
\label{eq:H_planet}
\end{align}
Eq.\,\eqref{eq:H_planet} is analogous to the rate of change of semi-major axis due to the external specific torque, $\Gamma$, done by the disk \citep[see e.g.,][]{BenitezLlambay2024}
\begin{equation}
\dot{a}_{\rm p} = \frac{2\Gamma}{v_{\rm p}}\,.
\end{equation}

\subsection{Initial and boundary conditions}

Initial and boundary conditions are crucial for solving the equations in comoving coordinates. While initial conditions are handled in a straightforward manner, the treatment of boundaries is a key differentiator of the comoving frame. Models defined in the fixed inertial frame can be adapted easily by transforming the inertial variables into their comoving counterparts. For instance, this means an initial power-law density profile defined in the inertial frame transforms directly into a new power-law profile within the comoving coordinate system. 
In the comoving frame, the computational mesh is fixed relative to the planet, meaning the boundaries are stationary within the comoving coordinates. From the perspective of the inertial frame, however, these boundaries expand or contract as the planet migrates, creating a moving-boundary problem analogous to a standard Stefan problem. This introduces a key challenge: for simulations where the global disk structure is important, the comoving boundaries must be informed by the large-scale evolution. 
This can be achieved, for example, by coupling the simulation to a global background model that feeds information to the comoving boundaries as needed. A similar approach was successfully used to simulate the global viscous disk evolution for the remapping method \cite{BenitezLlambay2016b}. 
Finally, if wave-damping zones are implemented at the grid edges, their damping recipes must also be written in comoving coordinates and variables to ensure they indeed force the system to the desired solution.

\section{Comoving framework in practice}
\label{sec:implementation}

\subsection{Theoretical Speed-up in a Comoving Frame}

We now derive a first-order estimate for the computational advantage of employing a comoving coordinate system over a standard, stationary physical frame for hydrodynamical simulations of inwardly migrating planets.

Consider two computational domains, a stationary physical grid and a comoving grid that follows the planet. 
Let the number of radial zones in the physical and comoving grids be $n_{\rm f}$ and $n_{\rm c}$, respectively. The physical grid has fixed inner and outer boundaries at $r=\mathcal{R}^{-}$ and $r=\mathcal{R}^{+}$, while the comoving grid has boundaries defined relative to the planet's position $a_{\rm p}$, at $r'=\mathcal{R}'^{-}$ and $r' = \mathcal{R}'^{+}$, such that its physical boundaries are at $r=a_{\rm p}(t)\mathcal{R}'^{-}$ and $r=a_{\rm p}(t)\mathcal{R}'^{+}$.

By employing algorithms such as FARGO \citep{Masset2000} or RAM \citep{BenitezLlambay2023}, the Courant-Friedrichs-Lewy (CFL) time-step constraint imposed by Keplerian shear can be circumvented. Consequently, the simulation time step, $\Delta t$, is primarily determined by the sound-crossing time across a grid cell at the inner computational boundary, $r_{\rm ib}$ \cite[see e.g.,][]{BenitezLlambay2023}
\begin{equation}
\label{eq:delta_t}
    \Delta t = C_{\rm CFL} \left.\frac{\Delta r}{c_{\rm s}}\right|_{r = r_{\rm ib}}\,,
\end{equation}
where $C_{\rm CFL}<1$ is the Courant number. This condition assumes that the grid cells are approximately square, i.e., the azimuthal spacing $r\Delta \varphi$ is comparable to the radial spacing $\Delta r$.

We adopt two common model assumptions. First, we use a logarithmic radial mesh, where the cell width $\Delta r$ is proportional to the radius $r$, such that $\Delta r = k r$, where $k$ is a constant defining the mesh resolution. Second, for a non-flared disk model, the sound speed $c_{\rm s}$ follows the power law $c_{\rm s}(r) = c_{{\rm s},0} (r/r_0)^{-1/2}$, with $c_{{\rm s},0} = h_0 \Omega_0 r_0$. Substituting these into Eq.\,\eqref{eq:delta_t}, the time step is given by
\begin{equation}
    \Delta t = \frac{C_{\rm CFL} k}{h_0 \Omega_0} \left(\frac{r_{\rm ib}}{r_0}\right)^{3/2}\,.
\end{equation}
We will absorb $C_{\rm CFL}$ into the definition of computational cost below and proceed with the proportionality. We compare the computational cost for two distinct migration scenarios: a constant physical migration rate and a constant comoving migration rate.

\subsubsection{Case I: Constant $\dot{a}_{\rm p}$}

We first assume a planet migrating inwards at a constant physical rate $\dot{a}_{\rm p} < 0$ from an initial position $a_{\rm p,i}$ to a final position $a_{\rm p,f}$ over a time $t_m = (a_{\rm p,f} - a_{\rm p,i}) / \dot{a}_{\rm p}$.

In the inertial fixed frame, the inner boundary $r_{\rm ib} = \mathcal{R}^{-}$ is fixed. Therefore, the time step $\Delta t_{\rm fix}$ is constant throughout the simulation. The total computational cost, $p_{\rm fix}$, defined as the total number of floating-point operations, is proportional to the number of cells, $n_{\rm f}$, multiplied by the number of time steps $N_{\rm fix} = t_m / \Delta t_{\rm fix}$
\begin{equation}
 p_{\rm fix} \propto n_{\rm f} N_{\rm fix} = h_0 \frac{n_{\rm f}}{k_{\rm f}}  \left(\frac{\Omega_0 r_0}{\dot{a}_{\rm p}}\right) \left(\frac{a_{\rm p,f}}{r_0}-\frac{a_{\rm p,i}}{r_0}\right) \left(\frac{\mathcal{R}^{-}}{r_0}\right)^{-3/2}\,.
\end{equation}
Here, $k_{\rm f}$ is the resolution constant for the fixed grid. 

In the comoving frame, the grid shrinks along with the planet's orbit. The inner boundary is located at $r_{\rm ib}(t) = a_{\rm p}\mathcal{R}'^{-}$. This causes the time step $\Delta t_{\rm com}(t)$ to decrease as the simulation progresses 
\begin{equation}
    \Delta t_{\rm com}(t) = \frac{k_{\rm c}}{h_0 \Omega_0} \left(\frac{a_{\rm p} \mathcal{R}'^{-}}{r_0}\right)^{3/2}\,.
\end{equation}
The total number of steps, $N_{\rm com}$, must be found by integrating the time-step duration over the total migration time
\begin{equation}
\label{eq:N_com_2}
    N_{\rm com} = \int_0^{t_m} \frac{dt}{\Delta t_{\rm com}} = \int_{a_{\rm p,i}}^{a_{\rm p,f}} \frac{da_{\rm p}}{\dot{a}_{\rm p} \Delta t_{\rm com}}\,.
\end{equation}
The computational cost, $p_{\rm com}$, is proportional to $n_{\rm c} N_{\rm com}$. After evaluating the integral \eqref{eq:N_com_2}, we find
\begin{equation}
    p_{\rm com} \propto n_c N_{\rm com} = 2h_0 \frac{n_{\rm c}}{k_{\rm c}} \left(\frac{\Omega_0 r_0}{\dot{a}_{\rm p}}\right) \left(\sqrt{\frac{r_0}{a_{\rm p,i}}} - \sqrt{\frac{r_0}{a_{\rm p,f}}}\right) \left(\mathcal{R}'^{-}\right)^{-3/2}\,.
\end{equation}

\subsubsection{Case II: Constant $\mathcal{H}$}

We now consider the migration regime where $\mathcal{H} = \dot{a}_{\rm p}/v_{\rm p}$ is constant as in the benchmark test presented in Section \ref{sec:benchmark}.
In this case, the total migration time is 
\begin{equation}
    t_m = \frac{1}{\mathcal{H}} \int_{a_{\rm p,i}}^{a_{\rm p,f}} \frac{da_{\rm p}}{a_{\rm p}\Omega_{\rm p}} = \frac{2}{3\mathcal{H}\Omega_0}\left(\left(\frac{a_{\rm p,f}}{r_0}\right)^{3/2}-\left(\frac{a_{\rm p,i}}{r_0}\right)^{3/2}\right) \,,
\end{equation}
and the computational cost, $p_{\rm fix}$ is
\begin{equation}
 p_{\rm fix} \propto \frac{2}{3} h_0 \frac{n_{\rm f}}{k_{\rm f}} \left(\frac{1}{\mathcal{H}}\right) \left(\left(\frac{a_{\rm p,f}}{r_0}\right)^{3/2}-\left(\frac{a_{\rm p,i}}{r_0}\right)^{3/2}\right)\left(\frac{\mathcal{R}^{-}}{r_0}\right)^{-3/2}\,.
\end{equation}
As in the previous case, the number of steps $N_{\rm com}$ is found through integration, yielding a comoving cost
\begin{equation}
    p_{\rm com} \propto h_0 \frac{n_{\rm c}}{k_{\rm c}} \left(\frac{1}{\mathcal{H}}\right) \log\left(\frac{a_{\rm p,f}}{a_{\rm p,i}}\right) \left(\mathcal{R}'^{-}\right)^{-3/2}\,.
\end{equation}

\begin{figure}
    \centering
    \includegraphics[]{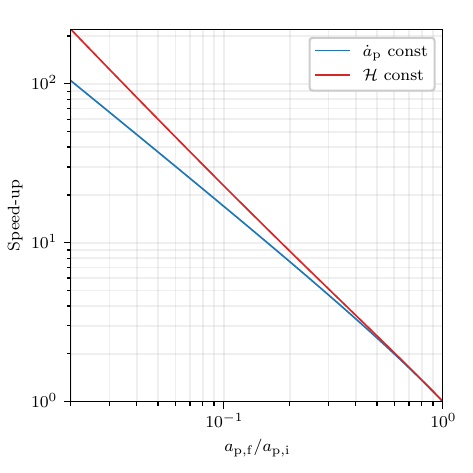}
    \caption{Theoretical speed-up of the comoving framework (inverse of Eq.\,\ref{eq:relative_cost_gen}) relative to a fixed-grid simulation, plotted as a function of the migration extent $a_{p,f}$ relative to the planet's  initial semi-major axis, $a_{p,i}$. The speed-up is calculated for models with constant $\dot{a}$ (blue line) and constant $\mathcal{H}$ (red line). The latter corresponds to the steady-state solution found in our benchmark test (section \ref{sec:benchmark}), where the physical migration rate scales as $\dot{a}_p \propto a_p^{-1/2}$. The computational advantage of the comoving frame grows significantly as the planet migrates further inwards. The speed-up is particularly pronounced for the case of constant $\mathcal{H}$ because the planet spends more physical time at larger radii, where the fixed-grid simulation is severely penalized by the small time step imposed at its inner boundary.}
    \label{fig:speed-up}
\end{figure}

\subsubsection{Relative Computational Cost}

We compare the costs by taking the ratio $p_{\rm com}/p_{\rm fix}$. 
To ensure a fair comparison, both simulations must have the same effective resolution. We define this by requiring a constant number of cells per pressure scale height, $N_H$. Assuming a constant disk aspect ratio $H/r = h_0$, the resolution per scale height is $N_H = H/\Delta r = h_0/k$. Thus, for a fixed $N_H$, the constant $k = h_0/N_H$ is the same for both setups, so $k_{\rm f} = k_{\rm c}$.

The number of radial cells $n$ is related to the grid's inner ($r_{\rm ib}$) and outer ($r_{\rm ob}$) boundaries and the resolution constant $k$ by $r_{\rm ob} = r_{\rm ib}(1+k)^{n-1}$. For $n \gg 1$, this gives
\begin{equation}
\label{eq:n}
n \approx \frac{\log(r_{\rm ob}/r_{\rm ib})}{\log(1+k)}\,.
\end{equation}
Therefore, the ratio of cell numbers is $n_{\rm c}/n_{\rm f} \approx \log(\mathcal{R}'^{+}/\mathcal{R}'^{-}) / \log(\mathcal{R}^{+}/\mathcal{R}^{-})$.

After substituting these relations, we arrive at the final expression for the relative cost for both cases
\begin{equation}
\label{eq:relative_cost_gen}
    \frac{p_{\rm com}}{p_{\rm fix}} = \mathcal{F} \times \frac{\log\left(\mathcal{R}'^{+}/\mathcal{R}'^{-}\right)}{\log\left(\mathcal{R}^{+}/\mathcal{R}^{-}\right)} \left(\frac{ \mathcal{R}'^{-}}{\mathcal{R}^{-}}\right)^{-3/2}\,,
\end{equation}
where the factor $\mathcal{F}$ depends on the migration model
\begin{equation}
    \mathcal{F} = 
    \begin{cases} 
      \displaystyle{2 \frac{a_{\rm p,i}^{-1/2} - a_{\rm p,f}^{-1/2}}{ a_{\rm p,f}-a_{\rm p,i} }} & \text{if } \dot{a}_{\rm p} = \text{const} \\
      \displaystyle{\frac{3}{2} \frac{\log(a_{\rm p,f}/a_{\rm p,i})}{ a_{\rm p,f}^{3/2}-a_{\rm p,i}^{3/2} }} & \text{if } \mathcal{H} = \text{const}
   \end{cases}\,.
\end{equation}
The inverse of Eq.\,\eqref{eq:relative_cost_gen} is the computational speed-up of the comoving method.
We illustrate this speed-up in Figure~\ref{fig:speed-up} as a function of the migration extent $a_{p,f}$ relative to the planet's initial semi-major axis, $a_{p,i}$. For that, we consider an inward migration from $a_{\rm p,i}$ to $a_{\rm p,f}$, a fixed grid covering the full path ($\mathcal{R}^{-} = 3^{-2/3} a_{\rm p,f}$ to $\mathcal{R}^{+} = 3^{2/3} a_{\rm p,i}$), and a comoving grid centered on the planet ($\mathcal{R}'^{-} = 3^{-2/3}$ to $\mathcal{R}'^{+} = 3^{2/3}$).  

This figure shows that the efficiency of the comoving frame grows as the planet migrates inwards. The speed-up is more pronounced for the case of constant $\mathcal{H}$ because the planet spends more physical time far from the star, where the fixed-grid simulation is more expensive due to the time step imposed at its inner boundary.
Our result implies that for planets that migrate over a decade in radius, the comoving simulation is $\mathcal{O}(10)$ faster than the equivalent simulation on a fixed grid for both migration models. This speed-up factor is independent of the simulation's dimensionality (1D, 2D, or 3D), as the number of cells in the non-radial directions would be identical in both frames and thus cancel out in the cost ratio. We emphasize that the final speed-up obtained in a practical situation will be dependent on the migration track followed by the planet.

\subsection{Implementation in FARGO3D}

To validate the comoving framework presented in Section \ref{sec:comoving_transformation}, we have implemented it into the hydrodynamic code FARGO3D\footnote{In Appendix \ref{sec:riemann_solver} we describe a possible implementation of the comoving framework on codes based on Riemann Solvers.} \citep{BenitezLlambay2016a}.
This implementation primarily involves the calculation and inclusion of the new source term, {\bf S}, within the momentum equation, with minimal changes to the standard source code. 
As described in the preceding section, the planet's equation of motion is solved in rest frame; therefore, transformations between the rest frame and the comoving frame are required to connect the fluid and planetary equations. 
Our implementation of the comoving framework in FARGO3D is described in the following algorithm and summarized in Table \ref{alg:comoving}:
\begin{enumerate}
    \item Set the initial radial distance, $a_{p}$, and velocity, $\dot{a}_{p}$, for the comoving frame. For a planet initially at rest, $\dot{a}_{p} = 0$. Subsequently, calculate the angular frequency, $\Omega_{p}$, and the dimensionless scaling rate, $\mathcal{H}$, for the initial state and transform the initial rest-frame model into the comoving frame using the comoving variables.

    \item At each time step, calculate the gravitational potential of the planet. Next, update the planet's position in rest frame and use this new position to compute the advanced values for $a_{\rm p}$, $\dot{a}_{\rm p}$, $\Omega_{\rm p}$, and $\dot{\mathcal{H}}$. With these values updated, calculate the source term ${\bf S}$, which accounts for the fictitious accelerations introduced by the comoving frame, and add it to the FARGO3D source step \citep[see][]{BenitezLlambay2016a}. The hydrodynamic equations are then solved using a standard solver (in comoving coordinates) to complete the full time step.

    \item Repeat the previous step until the simulation reaches its designated end time, $t_{end}$ or $t'_{\rm end}$.
\end{enumerate}

\begin{table}[h!]
    \centering
    \caption{Comoving framework in FARGO3D}
    \label{alg:comoving}
    \begin{tabular}{r l}
        \hline\hline
        \multicolumn{2}{l}{\textbf{Comoving Algorithm}} \\
        \hline
        1. & Set initial $a_p$, $\dot{a}_p$. \\
        2. & Calculate initial $\Omega_p$ and $\mathcal{H}$. \\
        3. & Transform initial rest-frame model to the comoving frame using equations \eqref{eq:comoving_density}-\eqref{eq:comoving_velocity}. \\
        4. & \textbf{for} $t'=0$ to $t'_{\rm end}$ with step $\Delta t'$ \textbf{do} \\
        5. & \hspace*{1.5em} Calculate the planet's potential using Eq.\,\eqref{eq:comoving_potential}. \\
        6. & \hspace*{1.5em} Update the planet's position in the rest frame. \\
        7. & \hspace*{1.5em} Use the new position to calculate advanced values for $a_{\rm p}$, $\dot{a}_{\rm p}$, $\Omega_{\rm p}$, $\mathcal{H}$, and $\dot{\mathcal{H}}$. \\
        8. & \hspace*{1.5em} Calculate the source term $\mathbf{S}$ and add it to the so-called source step. \\
        9. & \hspace*{1.5em} Solve the hydrodynamic equations as if they were in the rest frame with the source term $\mathbf{S}$ added. \\
        10.& \textbf{end for} \\
        \hline\hline
    \end{tabular}
\end{table}

\subsection{Benchmark Test and Algorithm Validation}
\label{sec:benchmark}

To validate our numerical implementation and illustrate the advantages of the comoving framework, we present a benchmark test. This test is designed to compare the results of a planet migration simulation performed in the comoving frame against an equivalent simulation in the standard inertial frame.

The setup consists of a planet of mass $m_{\rm p}$ on an initial circular orbit with a semi-major axis $a_{\rm p}(t=0) = a_{\rm p,0}$. The planet's gravitational potential is modeled as a Plummer potential with a smoothing length $\epsilon$, and it is free to migrate due to the gravitational force exerted by a two-dimensional, non-self-gravitating gaseous disk.

We model the protoplanetary disk with a surface density profile given by the power-law
\begin{equation}
    \Sigma(r) = \Sigma_0 \left(\frac{r}{r_0}\right)^{-\sigma},
\end{equation}
and a locally isothermal equation of state, where the pressure $P$ is related to the surface density by $P = c_{\rm s}^2 \Sigma$. The sound speed is defined as $c_{\rm s} = h_0 v_{\rm K}$, where $h_0$ is the constant disk aspect ratio and $v_{\rm K}$ is the Keplerian velocity. We adopt a standard $\alpha$-disk model for viscosity, with a constant kinematic viscosity parameter $\alpha_{\nu}$. The initial velocity field corresponds to an approximate steady-state solution for a viscous accretion disk
\begin{align}
    v_r &= -\frac{3 \alpha_{\nu} h_0 }{2} c_{\rm s}\,, \\
    v_\varphi &= \left[1-h^2 (1+\sigma)\right]^{1/2} v_{\rm k}\,.
\end{align}

In the 2D comoving frame ($d=2$), the initial surface density and pressure are given by
\begin{equation}
\label{eq:example_sigma}
    \Sigma'(r') = \left(\frac{a_{\rm p}^{2}\Sigma_0}{M_\star}\right) \left(\frac{a_{\rm p}}{r_0}\right)^{-\sigma} r'^{-\sigma}\,,\\
\end{equation}
\begin{equation}
    P' = c_{\rm s}'^2 \Sigma'\,,
\end{equation}
with a comoving sound speed $c_{\rm s}'(r') = h_0 r'^{-1/2}$. The corresponding comoving velocities are
\begin{align}
    u'_r &= -\frac{3\alpha_{\nu} h_0^2}{2} r'^{-1/2}  - r' \mathcal{H}\,,\\
    u'_\varphi &= \left[1-h^2 (1+\sigma)\right]^{1/2} r'^{-1/2} \,.
\end{align}
Eq.\,\eqref{eq:example_sigma} shows that if the density slope is $\sigma = 2$, the term $(a_{\rm p}/r_0)^{-2}$ cancels with the scaling factor $a_{\rm p}^2$, resulting in a time-independent comoving surface density profile. Our test is designed to verify that our simulation correctly recovers this steady-state solution, which manifests as an exponential decay in comoving coordinates. 

\subsubsection{Numerical Setup}

\begin{figure}[t!]
    \centering
    \includegraphics[]{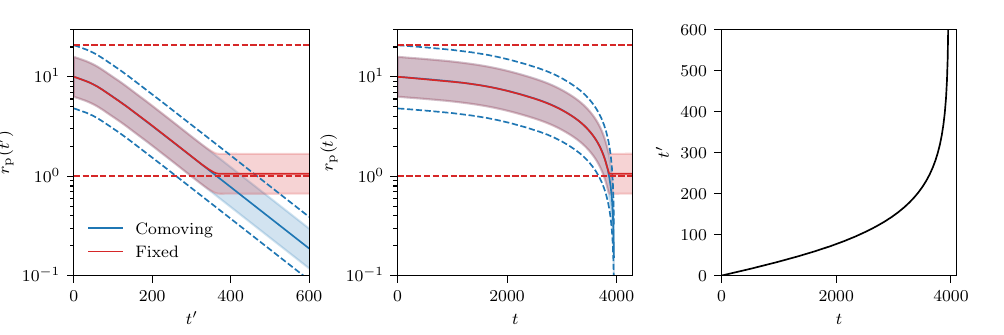}
        \caption{Results of the benchmark test comparing the planet migration trajectories from the comoving (blue) and inertial frame (red) simulations. In all panels, solid lines represent the planet's radial position, dashed lines show the boundaries of the respective computational domains, and the shaded regions indicate the co-orbital ring within which gravitational torques are calculated. The left panel shows the planet's radial position $r_{\rm p}$ versus comoving time $t'$. The linear decay on the log-linear plot indicates a constant migration rate in the comoving frame, consistent with the expected steady-state solution. The middle panel shows the planet's radial position versus time $t$. The agreement between the two methods is excellent until the planet in the inertial frame simulation approaches the inner grid boundary, causing its migration to stall. The right panel shows the non-linear transformation between time and comoving time given by Eq.\,\eqref{eq:tprime}, which depends on the planet trajectory.}
    \label{fig:benchmark}
\end{figure}

The parameters for this test are $q = 3\times 10^{-4}$, $\Sigma_0 = 4\times 10^{-3}$, $\sigma=2$, $h_0 = 5\times10^{-2}$, $\epsilon=0.6 h_0 r_{\rm p}$, and $\alpha_{\nu} = 3\times 10^{-3}$. These values are chosen to produce rapid planet migration, providing a stringent test of our framework. The planet is initially placed at $a_{\rm p,0} = 10$ with $\dot{a}_{\rm p}(0) = 0$, which implies $\mathcal{H}(0)=0$. Units are chosen such that the gravitational constant $G$, the central mass $M_\star$, and the reference radius $r_0$ are all unity.

We employ a polar logarithmically spaced radial grid and a uniform azimuthal grid. In the comoving frame, the computational domain spans $r' \in [3^{-2/3}, 3^{2/3}]$. For the inertial frame simulation, the domain is set to $r \in [1, 10\times3^{2/3}]$ to accommodate the planet's inward migration. The resolution is set to approximately 16 cells per local scale height, which results in $n_r=469$ radial cells for the comoving grid and $n_r=973$ for the larger static grid. To obtain cell aspect ratios close to unity, we use $n_\varphi = 2011$ azimuthal cells spanning $\varphi \in [-\pi, \pi]$.

To ensure a robust validation of our comoving frame implementation, we must account for the known
sensitivity of torque calculations to the position of grid boundaries \citep{BenitezLlambay2016b}. As a planet in an inertial frame moves closer to the inner boundary, the computed torque can change, causing its trajectory to diverge from that of a comoving frame simulation where the planet remains centered. To perform a precise, one-to-one comparison and isolate potential implementation issues, it is crucial that the gravitational torque on the planet is computed from an identical region of the disk in both simulations. We achieve this by restricting the disk's gravitational force calculation to a co-orbital ring centered on the planet's instantaneous position. The ring's radial extent is chosen to be wide enough to contain the 3:2 mean-motion resonances, as the planet migrates. This constraint, imposed solely for the rigor of this validation test and not as a general requirement of the framework, ensures that any observed discrepancies between the two simulations can be attributed directly to the numerical framework rather than to differing boundary effects or implementation issues.
In addition, our force calculation includes a necessary correction because disk self-gravity is not included in our simulations. Without this correction, the planet and the gas would orbit a different effective potential, as the planet is affected by the disk's gravity but the disk does not feel its own. This inconsistency leads to a spurious shift in the locations of Lindblad resonances. To remedy this, we adopt the method described in \cite{BenitezLlambay2016b}, which consists of subtracting the azimuthally-averaged density profile before computing the gravitational force exerted by the disk on the planet. This procedure does not alter the net torque but corrects the planet's orbital frequency, making it dynamically consistent with the gas rotation profile \citep{Baruteau2008}.
In both simulations, standard wave-damping zones are implemented near the grid boundaries, far from the torque calculation ring, to prevent wave reflection. 

\subsubsection{Results}

\begin{figure}[t!]
    \centering
    \includegraphics[]{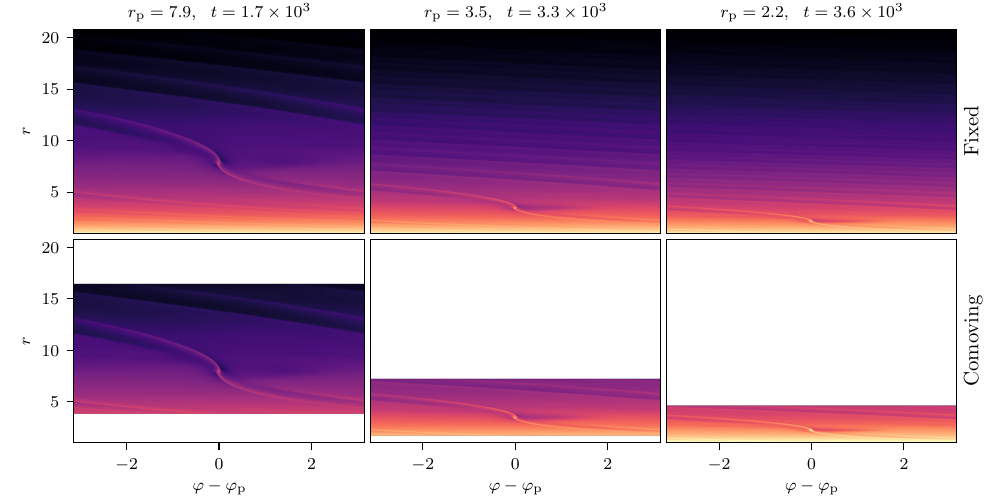}
    \caption{Snapshots of the gas surface density, $\Sigma(r)$, in physical coordinates, comparing the fixed inertial frame simulation (top row) with the comoving frame simulation (bottom row). The surface density is plotted against the radius, $r$, and the azimuthal angle relative to the planet, $\varphi - \varphi_{\rm p}$. For direct comparison, the comoving simulation is plotted on the same physical domain as the fixed-grid one. Columns correspond to three different times. The surface density is shown using a  logarithmic scale with a linear color map over the range -5.0 (darker color) to -2.5 (lighter color). The figure illustrates the contraction of the comoving frame's domain in physical space as it follows the planet inward.}
    \label{fig:fixed_frame}
\end{figure}

\begin{figure}[t!]
    \centering
    \includegraphics[]{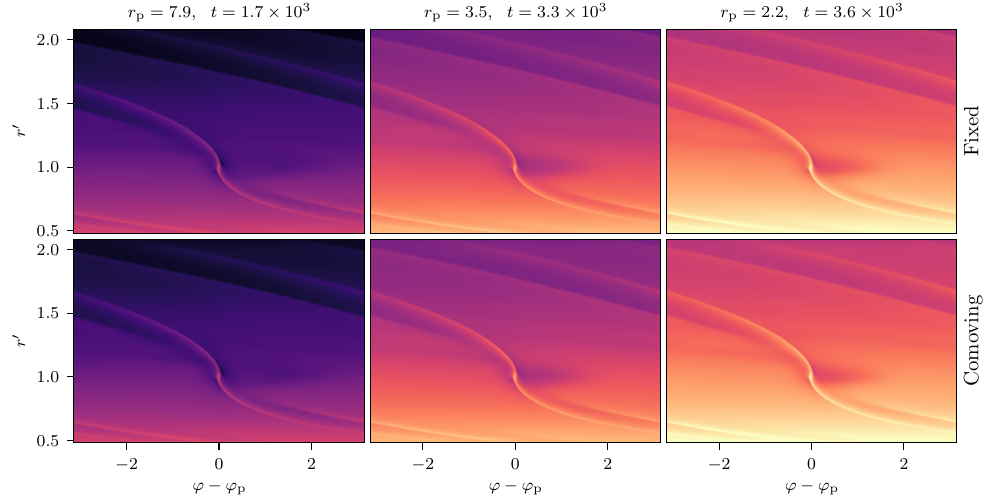}
    \caption{Snapshots of the gas surface density in comoving coordinates, comparing the fixed inertial frame (top row) and the comoving frame (bottom row) simulations at the same times as in Fig.~\ref{fig:fixed_frame}. The surface density is plotted against the comoving radius, $r'$, and the azimuthal angle relative to the planet, $\varphi - \varphi_{\rm p}$. The surface density is shown using a  logarithmic scale with a linear color map over the range -5.0 (darker color) to -2.5 (lighter color).
    The excellent agreement between the disk structures shows that the comoving framework accurately reproduces the local physics of planet-disk interaction of migrating planets. 
    As the planet migrates inward, the background surface density within the comoving frame correctly increases, consistent with the initial density profile.}
    \label{fig:comoving_frame}
\end{figure}

\begin{figure}[t!]
    \centering
    \includegraphics[]{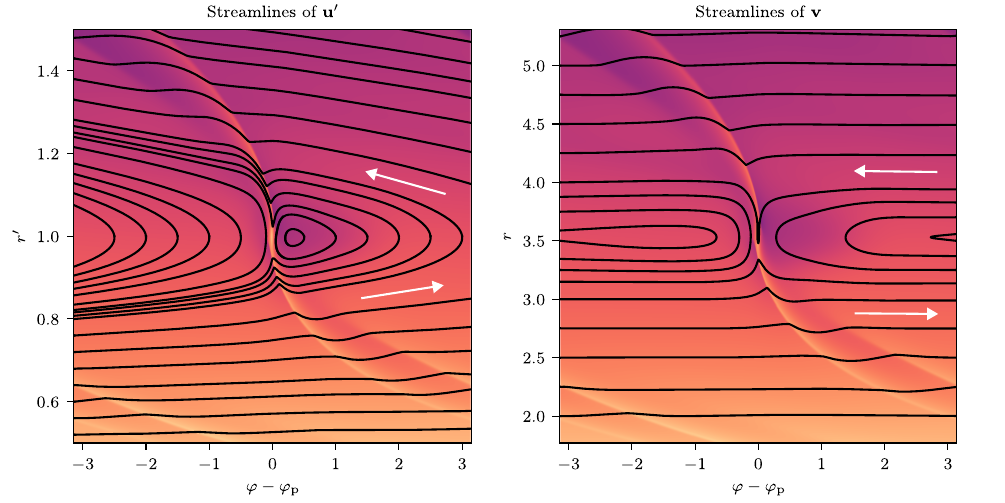}
    \caption{Comparison of the gas flow streamlines around a migrating planet, shown in the comoving frame from a comoving simulation (left) and the inertial frame from a fixed grid simulation (right) for a single time snapshot ($t=3.3\times10^3)$. The left panel displays the streamlines of the dimensionless comoving velocity, ${\bf u}'$ in comoving coordinates. The right panel shows the corresponding streamlines of the inertial velocity, ${\bf v}$, in physical coordinates. The white arrows in both panels point towards the instantaneous direction of the flow. Because the comoving frame allows the system to reach a steady state, the streamlines of ${\bf u}'$ also represent the true trajectories of fluid elements. In the time-dependent inertial frame, the streamlines of ${\bf v}$ are an instantaneous representation of the velocity field and do not correspond to particle paths. The background color map in both panels shows the gas surface density on a logarithmic scale following Figs.\,\ref{fig:fixed_frame} and \ref{fig:comoving_frame}.}
    \label{fig:streamlines}
\end{figure}

The results of our benchmark test, which directly compares a simulation in the comoving frame against an equivalent one in a standard inertial frame, are presented in Figures \ref{fig:benchmark}, \ref{fig:fixed_frame}, \ref{fig:comoving_frame}, and \ref{fig:streamlines}. These figures validate our implementation and showcase the key advantages of the comoving framework.
Figure~\ref{fig:benchmark} shows the planet's radial trajectory. The left panel shows the planet's radial position, $r_{\rm p}$, as a function of the comoving time, $t'$. The linear trend on this log-linear plot clearly shows that the planet reaches a steady-state migration with a constant migration rate in the comoving frame  consistent with the expected theoretical solution for our chosen parameters. The value measured from the simulation is $\mathcal{H}=6.3662$.
The middle panel displays the same migration but against time, $t$. The trajectories from the comoving (blue) and inertial (red) simulations show excellent agreement throughout the initial phase of the migration. However, the simulations diverge as the planet in the inertial frame approaches the inner boundary of its fixed computational grid. At this point, the torque calculation is compromised, and the planet's inward migration artificially stalls. In contrast, the planet in the comoving frame can continue its migration indefinitely. The right panel explicitly shows the non-linear transformation between the two time coordinates, which arises because the time step becomes progressively smaller (in units of the fixed frame) as the comoving grid shrinks to smaller radii.

Figures~\ref{fig:fixed_frame} and \ref{fig:comoving_frame} provide a direct visual comparison of the gas surface density, $\Sigma$, at three distinct time snapshots. Figure~\ref{fig:fixed_frame} displays the disk structure in physical coordinates. The top row shows the simulation in the fixed inertial frame, while the bottom row shows the result from the comoving frame, plotted over the same absolute radial range. This figure clearly illustrates how the comoving grid, which remains compact around the planet, contracts in physical space as it follows the planet's inward migration. However, the surface-density within the planet region is the same in both frames. 
Another view of the same data is presented in Figure~\ref{fig:comoving_frame}, where the same snapshots are shown in comoving coordinates. The excellent match between the inertial (top row) and comoving (bottom row) simulations demonstrates that our framework and implementation reproduces the results obtained with classical methods. The panels also show the surface density of the disk increasing as the planet moves to smaller radii, which is consistent with the initial power-law density profile ($\Sigma \propto r^{-2}$) when viewed on a shrinking coordinate system. 

Figure \ref{fig:streamlines} provides a direct comparison of the gas flow topology in the comoving and inertial frames. The left panel shows the streamlines of the dimensionless comoving velocity, ${\bf u}'$, in comoving coordinates. A key feature of our framework is that the system can reach a steady state in the comoving frame, with a constant migration rate. Consequently, these streamlines represent the true trajectories of fluid elements. In contrast, the right panel displays the instantaneous streamlines of the velocity field, ${\bf v}$, in the inertial frame. Because the planet is in constant radial motion in this frame, the flow is inherently time-dependent. Therefore, these streamlines only offer a snapshot of the velocity directions at a given moment and do not correspond to the actual paths of fluid particles, making it harder to analyze the disk dynamics from this perspective. This comparison highlights a conceptual advantage of the comoving frame, which transforms a complex, time-dependent flow into a steady-state problem where streamlines and trajectories become equivalent.

Finally, we report a speedup $\sim O(10)$ obtained for this particular configuration, which is consistent with the first-order estimation presented in the previous section.

Collectively, these results confirm that the comoving framework is a robust, accurate, and efficient method for simulating long-term planet migration.

\section{Discussion}
\label{sec:discussion}

The comoving framework introduced in this paper offers a robust and computationally efficient solution to the long-standing challenge of simulating planet migration self-consistently. Our benchmark test has validated its implementation in a hydrodynamical code, demonstrating its advantage over traditional fixed-grid methods. Beyond its clear utility as a numerical method, the framework also lays the groundwork for theoretical advances in the study of planet-disk interaction. Here, we discuss its broader implications and outline promising avenues for future research.

\subsection{Extension to Multi-Planet and Eccentric Systems}
\label{sec:multiplanet}

While we have focused on a single planet on a quasi-circular orbit, our framework is adaptable to more complex system architectures. For systems with multiple planets, several strategies can be employed. One straightforward approach is to center the comoving frame on a single planet and define a computational domain large enough to encompass the orbits of all other planets. Alternatively, the frame's scaling could be tied to a radius representing the system's average location, such as the semi-major axis of the planets' common center of mass or any other point in between the bodies. An advanced and potentially more efficient approach for widely separated planets would be to develop a multi-mesh system, where each planet resides in its own comoving grid, allowing for localized high resolution calculations of the disk response to each body as they migrate.

The comoving framework also presents a powerful tool for studying planets on eccentric orbits. By normalizing the spatial coordinates with the planet's instantaneous radial distance, $r_{\rm p}$, instead of its semi-major axis, $a_{\rm p}$, a planet on an eccentric path can be held at a fixed position within the computational grid. This capability is particularly advantageous when combined with mesh refinement techniques, such as nested meshes \citep{Velasco-Romero2024}, enabling ultra-high-resolution studies of the disk's response to eccentric perturbers without the need for an excessively large global domain. The synergy with advection algorithms designed for non-uniform meshes, like RAM \citep{BenitezLlambay2023}, further enhances the potential for detailed and highly efficient simulations.

\subsection{Long-Term Simulations}
\label{sec:longterm}

The computational efficiency of the comoving frame enables simulations that span secular timescales, where the global evolution of the protoplanetary disk becomes critically important. The local nature of our comoving grid necessitates a method to account for these large-scale changes. This can be achieved by coupling the local simulation to a global disk model that informs the boundary conditions of the comoving domain over time. For instance, a 1D or 2D axisymmetric disk evolution model \citep[e.g.][]{Robinson2024} could provide the necessary inflow/outflow conditions at the comoving grid's radial boundaries, a technique conceptually similar to that employed for the remapping method \cite{BenitezLlambay2016b}. This hybrid approach would capture both the detailed, local planet-disk interaction and the global evolution of the disk, leading to more realistic migration tracks.
Future work can extend this framework to include more complex physics, such as the two-fluid interactions of gas and dust or the treatment of flared disks with non-constant aspect ratios, using the formalisms we have outlined (see Sections \ref{sec:gas_dust} and \ref{sec:non_constant_h}).

\subsection{Alternative Comoving Transformations and Steady-State Solutions}
\label{sec:alternative_transformation}

The specific set of comoving variables defined in this work was chosen to render the continuity equation invariant. However, this is not the only possible choice. Different transformations can be designed to explore various physical regimes and study steady-state solutions for migrating planets. By choosing different comoving variables, one can analyze local migration phenomena as intermediate asymptotic solutions to the global disk evolution problem.

Furthermore, other coordinate transformations could be explored for local studies. As discussed previously in this paper, a transformation of the form $d{\bf r}' = d{\bf r}/H_{\rm p}$ is particularly interesting. For a non-flared disk, this naturally leads to a logarithmic mesh spacing, while for a flared disk, it defines a grid that strictly maintains a constant number of cells per scale height. Such a transformation, which combines a Galilean shift with a coordinate scaling, would be exceptionally useful for local studies that require constant resolution relative to the disk's vertical structure. This highlights the flexibility of the comoving approach not only as a numerical tool but as a theoretical framework to isolate and study the core physics of planet-disk interaction.

\subsection{Scope and Applicability of the comoving framework}

\paragraph{Coordinate Independence} The derivation of the comoving framework is expressed entirely in coordinate-free vector notation. Consequently, the validity of the method is independent of the specific geometry, as the additional source terms can be projected onto any coordinate system (see Section \ref{sec:comoving_transformation}).

\paragraph{Viscosity}
Although we implemented the comoving framework using an $\alpha$-disk model that renders the viscous term fully invariant, this is not strictly necessary. A general viscous law with zero bulk viscosity preserves the invariance of the stress tensor in comoving coordinates, even if the kinematic viscosity remains time-dependent. Consequently, simulations using general viscosity laws cannot reach a true steady state and must treat viscosity as a time-varying parameter. Similarly, this applies to an $\alpha$-disk model with a non-constant aspect ratio (see Section \ref{sec:viscosity})

\paragraph{Magnetohydrodynamics}
The induction equation in comoving coordinates is fully invariant in 3D. In lower dimensions, the framework remains applicable but requires the inclusion of additional source terms in the Induction equation to correct for the scaling of the magnetic field and maintain consistency with the Lorentz force (see Section \ref{sec:mhd}).

\paragraph{Thermodynamics}
The invariance of the energy equation in the comoving frame depends on the thermodynamic properties of the gas. For adiabatic simulations, the framework is strictly invariant only under certain conditions that depend on the dimensionality and $\gamma$. In a general case, a geometric source term must be included to account for the adiabatic cooling or heating induced by the expansion or contraction of the domain in comoving coordinates (see Section~\ref{sec:energy_equation}).

\paragraph{Disk geometry}
The current implementation in FARGO3D is optimized for disks with a constant aspect ratio. Nevertheless, the framework remains applicable to flared disks, provided the mesh adapts to resolve the pressure scale height as the planet migrates (see Section \ref{sec:alternative_transformation}). It is important to note that flared disks introduce a time dependence into the governing equations, precluding steady-state solutions within the comoving framework (see Section \ref{sec:non_constant_h}).

\paragraph{Multiple-planets}
The coordinate transformation is defined by a single scaling factor, making the framework best suited for tight planetary configurations where a common center of mass dominates the scaling. For multi-planet systems with widely separated orbits, the applicability of the method is maximized by extending it to a multi-mesh approach, where independent comoving grids track each body separately (see Section \ref{sec:multiplanet}).

\paragraph{Global evolution}
The comoving grid is local by definition, which creates a moving-boundary relative to the inertial frame. For short-term hydrodynamical studies, standard boundary conditions suffice. However, the scope of applicability for secular timescale simulations depends on the ability to account for the global disk evolution. Accurate long-term studies require coupling the comoving simulation to a global background model to inform the boundary conditions as the physical domain evolves (see Section \ref{sec:longterm}).

\section{Summary and Conclusions}
\label{sec:summary_conclusion}

Simulating the migration of planets within their natal protoplanetary disks presents a significant computational challenge. Standard numerical methods that employ a static grid are computationally expensive for long-term integrations, while alternative approaches that follow the planet can suffer from non-conservative numerical schemes. In this paper, we have introduced a novel comoving framework to address these limitations, which serves as both a practical numerical method and a theoretical tool to advance planet migration studies. Unlike remapping methods that rely on non-conservative interpolation of the solution at each timestep, our framework avoids this source of numerical diffusion entirely by transforming the hydrodynamic equations themselves.

Our method is based on a coordinate transformation that scales with the planet's evolving semi-major axis, effectively keeping the planet stationary with respect to the computational grid. This transformation modifies the standard hydrodynamic equations by introducing an additional source term that accounts for the inertial forces of the non-inertial, expanding/contracting reference frame. The framework evolves the equations of the complete star-planet-disk system self-consistently and it can be readily implemented in any hydrodynamics code with minimal modification.

We have provided a first-order theoretical estimate of the computational speed-up, finding that for a typical migration scenario, the comoving framework can be over an order of magnitude faster than an equivalent simulation on a fixed grid. We implemented our method in the FARGO3D code and validated it with a benchmark test of a migrating planet. The results show excellent agreement with the standard simulation on a fixed frame, successfully tracking the planet's migration long after the fixed-grid simulation fails due to the planet approaching the domain boundary.

The primary advantages of our comoving framework are

\paragraph{Computational Efficiency} It dramatically reduces the computational cost of simulating migrating planets by restricting the domain to the region of interest, making the cost comparable to that of simulations with fixed planets. This allows for a speed-up $>\mathcal{O}(10)$ with respect to standard fixed grid methods for large-scale migration.
\paragraph{Precision and Robustness} The framework is built on a direct transformation of the governing equations. By design, it avoids the mass and momentum interpolation required by methods like remapping, thereby preventing the introduction of unnecessary numerical diffusion and truncation errors.
\paragraph{Self-Consistency} It allows for a fully self-consistent treatment where the planet's orbit evolves in direct response to the gravitational force from the disk, whose structure is simultaneously shaped by the planet.

\medskip

Beyond its numerical efficiency, the framework provides a powerful conceptual simplification. It can recast the inherently time-dependent problem of a migrating planet interacting with the disk into a steady-state problem within the comoving reference frame. This transformation yields direct physical insight, as the resulting streamlines represent the true trajectories of fluid elements. This one-to-one correspondence between the flow field and particle paths is lost in the non-steady inertial frame, highlighting the comoving framework's strength as a theoretical tool for analyzing the fundamental dynamics of planet-disk interaction.

Furthermore, the dramatic reduction in numerical cost opens the door to incorporating more complex and realistic disk physics into self-consistent migration studies of single and multiple planets over extended timescales. 
Ultimately, the comoving framework is a powerful and versatile tool to address the core questions of planet formation. Its utility as a theoretical framework for analyzing steady-state solutions, combined with its power as a numerical method enabling hydrodynamical simulations of moving planets at a fraction of the traditional cost, represents a crucial advance. This work paves the way toward a grander goal: a comprehensive self-consistent theory of planet formation that can explain the origin of the diverse exoplanetary systems.

\begin{acknowledgments}
We are grateful to the referee for their constructive comments, which have significantly enhanced the quality of our manuscript. We also thank Alejandro Benítez-Llambay and Leonardo Krapp for their careful reading and critical suggestions, which greatly improved this paper.
P.~B.~L. acknowledges  support from ANID, QUIMAL fund ASTRO21-0039 and FONDECYT project 1231205. 
\end{acknowledgments}

\begin{contribution}
All authors contributed equally to the development of this paper.
\end{contribution}

\appendix 

\section{Comoving framework in Riemann Solvers}
\label{sec:riemann_solver}
Implementing the comoving framework in codes that rely on Riemann solvers requires minimal changes. 
Because the coordinate transformation absorbs the frame velocity into the definition of the fluid velocity ${\bf u}'$, the flux-conserving properties of the solvers are preserved. The implementation can be summarized in three steps:
\begin{enumerate}
    \item Source terms: add the comoving source terms \eqref{eq:source_term} to the right-hand side of the conservation laws. For conservative codes solving for momentum density, this term must be multiplied by the density $\rho'$. If solving the total energy equation, add the work done by the inertial forces ($\rho' {\bf u}' \cdot {\bf S}'$). Finally, the geometric source term shown in the R.H.S of Eq.\,\eqref{eq:energy_comoving} must be added to account for the expansion work of the frame.
    \item Spatial reconstruction: standard reconstruction applies directly to the comoving primitive variables because the grid coordinates are static within the comoving frame. The Riemann solver should compute fluxes using the reconstructed comoving velocities exactly as it would for a standard static mesh.
    \item Time-dependent terms: ensure that the scaling parameters ($a_{\rm p}$, $\dot{a}_{\rm p}$, $\Omega_p$, $\mathcal{H}$, $\dot{\mathcal{H}}$) are updated consistently with the time integration scheme. For multi-stage integrators, these parameters must be re-calculated at every substage to match the current simulation time $t'$.
\end{enumerate}

\end{document}